\documentclass[sigconf,screen,nonacm]{acmart}

\AtBeginDocument{%
  }

\renewcommand\footnotetextcopyrightpermission[1]{}
\usepackage{multirow} %ZZX
\usepackage{tabularx}
\begin{document}

\title{HoliDubber: Holistic Video Dubbing for Complex Acoustic Scenes via Text-Guided Audio Synthesis}


\author{Wenhao Guan}
\affiliation{
  \institution{Xiamen University, \\Shanghai Innovation Institute}
  \city{Xiamen}
  \country{China}
}

\author{Yifan Duan}
\affiliation{
  \institution{Shanghai Jiao Tong University}
  \city{Shanghai}
  \country{China}
}

\author{Junxi Liu}
\affiliation{%
  \institution{Shanghai Jiao Tong University}
  \city{Shanghai}
  \country{China}
}

\author{Yu Gu}
\affiliation{%
  \institution{Joy Future Academy}
  \city{Beijing}
  \country{China}
}

\author{Feng Dang}
\affiliation{
  \institution{Joy Future Academy}
  \city{Beijing}
  \country{China}
}

\author{Kaidi Wang}
\affiliation{
  \institution{Xiamen University}
  \city{Xiamen}
  \country{China}
}
\author{Qingyang Hong}
\authornote{Corresponding author.}
\affiliation{
  \institution{Xiamen University}
  \city{Xiamen}
  \country{China}
}
\author{Lin Li}
\authornotemark[1]
\affiliation{
  \institution{Xiamen University}
  \city{Xiamen}
  \country{China}
}

\author{Xie Chen}
\authornotemark[1]
\affiliation{
  \institution{Shanghai Jiao Tong University, \\Shanghai Innovation Institute}
  \city{Shanghai}
  \country{China}
}

\renewcommand{\shortauthors}{Wenhao Guan et al.}

\begin{abstract}
  Video dubbing is a cornerstone of multimedia content creation, aiming to synthesize synchronized acoustic sequences for visual streams. While Text-to-Speech (TTS) and Text-to-Audio (TTA) generation have each achieved remarkable progress, existing dubbing systems remain confined to isolated speech synthesis without incorporating sound effects and ambient audio, forcing practitioners to rely on fragmented workflows and laborious manual post-mixing.
  To address this limitation, we present HoliDubber, a holistic video dubbing framework that moves beyond speech-only generation by enabling the joint synthesis of speech and sound effects from a single text prompt. Specifically, HoliDubber adopts a patch-based autoregressive diffusion transformer architecture, where a causal language model autoregressively models aggregated patch embeddings to capture global temporal structure, and a Diffusion Transformer decoder generates high-fidelity continuous tokens within each patch, following a divide-and-conquer strategy. To achieve cross-modal alignment, visual features are encoded into patch-level representations and fused with audio patches via cross-attention, enabling the model to ground speech generation in the speaker's visual articulation dynamics. In addition, we introduce HoliDub-Bench, a benchmark curated from established datasets with synchronized video-text-audio triplets designed for holistic dubbing evaluation.
  Extensive experiments demonstrate that HoliDubber significantly outperforms existing methods across multiple benchmarks in speech quality, synchronization, and speaker similarity. Furthermore, results on HoliDub-Bench validate the effectiveness of joint speech-and-sound generation, establishing a new paradigm for holistic video dubbing in complex acoustic scenes. \footnote{The demo page of the project is \url{https://holidubber.github.io}}
\end{abstract}


\begin{CCSXML}
<ccs2012>
   <concept>
       <concept_id>10002951.10003227.10003251.10003256</concept_id>
       <concept_desc>Information systems~Multimedia content creation</concept_desc>
       <concept_significance>500</concept_significance>
       </concept>
 </ccs2012>
\end{CCSXML}

\ccsdesc[500]{Information systems~Multimedia content creation}


\keywords{Video Dubbing, Text-to-Speech, Text-to-Audio, Autoregressive Diffusion Transformer, Audio-Visual Synchronization}


\maketitle

\section{Introduction}

Video dubbing, the task of synthesizing synchronized audio tracks for visual content, has emerged as a fundamental technology in multimedia content creation. Traditional dubbing systems, often referred to as Visual Voice Cloning \cite{chen2022v2c}, aim to convert a given script into speech that matches a target voice while maintaining temporal synchronization with the character's lip movements and emotional coherence with their facial expressions. This technology has broad applications spanning film production, game development, digital media localization, and personalized speech generation. 

Existing approaches to video dubbing can be broadly organized into two research paradigms.
The first paradigm centers on audio-visual alignment, seeking to establish fine-grained synchronization between generated speech and the speaker's lip movements \cite{hu2021neuraldubber,cong2024styledubber}. 
While these methods achieve notable progress in lip-sync accuracy, they operate under a closed-set speaker assumption and lack the ability to generalize to unseen speakers in a zero-shot manner. 

The second paradigm targets zero-shot dubbing with a broader focus on speech quality, speaker similarity, and emotional expressiveness. 
Early efforts \cite{zhang2024speaker2dubber,cong2025emodubber,zhang2025produbber} in this direction build upon lightweight architectures such as FastSpeech-based backbones \cite{ren2019fastspeech}. 
More recently, the integration of Multimodal Large Language Models (MLLMs) has opened a new direction for video dubbing, enabling richer cross-modal fusion and more controllable speech generation through natural language instructions \cite{zheng2025deepdubber, zhang2026instructdubber, liu2026funcineforge}.

Despite significant progress, existing approaches remain confined to speech-only synthesis, treating dubbing as a purely linguistic task. However, the acoustic experience extends far beyond spoken dialogue in real-world video scenes.  Ambient sounds and sound effects all contribute to an immersive viewing experience. This gap between speech-centric dubbing and the holistic acoustic demands of real-world content forces practitioners to rely on fragmented pipelines, where speech synthesis, sound effect design, and audio mixing are handled by separate tools with laborious manual post-production. The challenge of generating coherent, multi-modal audio that simultaneously encompasses speech and sound effects, while maintaining fine-grained alignment with visual performance, remains largely unexplored.

In a related line of research, recent video-to-soundtrack methods \cite{tian2025dualdub, zhang2025deepaudio,cheng2025vssflow} have explored generating both speech and environmental sounds for video content. However, these approaches typically adopt a decoupled dual-head design that treats speech and audio as independent generation tasks, overlooking their natural entanglement in real-world scenes. 
Moreover, their reliance on visual signals as the primary generative condition limits the user's ability to explicitly control audio content through natural language.

\begin{figure}[!t]
  \centering
  \vspace{-9pt}
  \includegraphics[width=0.86\linewidth]{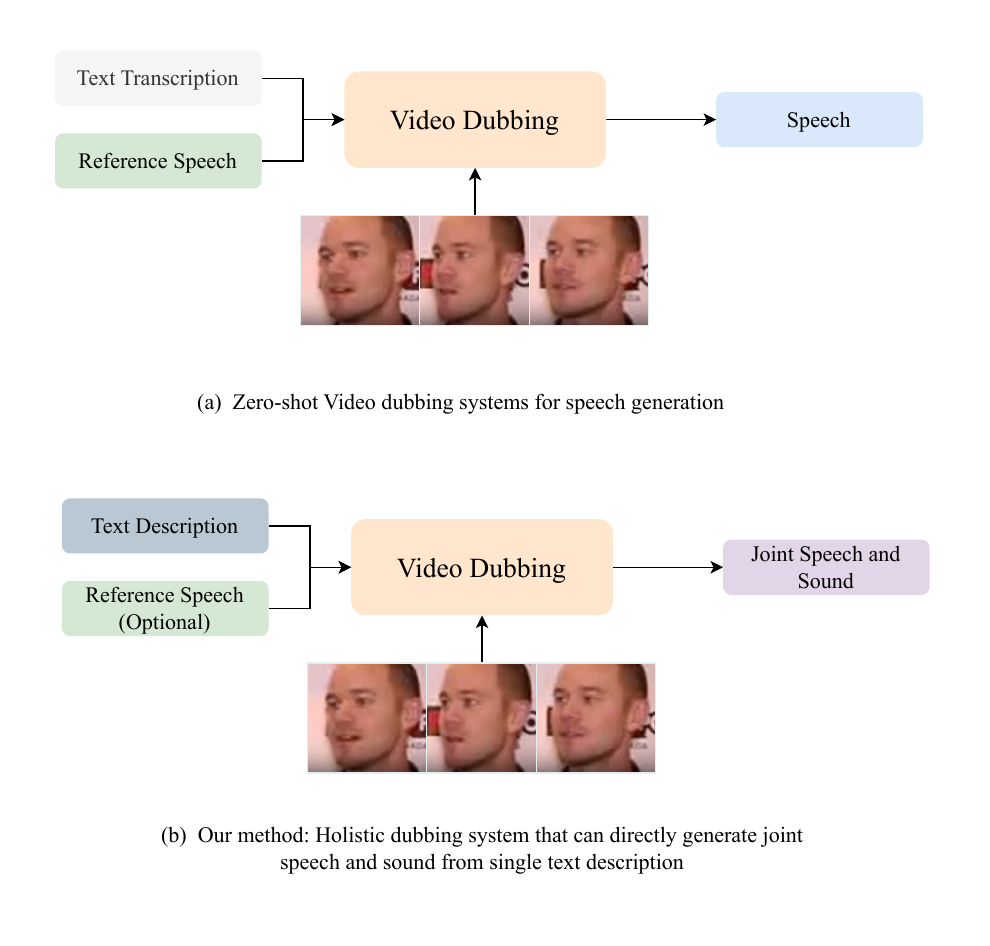}
  \caption{
  % Overview of previous video dubbing systems and our proposed HoliDubber system. 
  (a). Zero-shot video dubbing system for speech-only generation. (b). Unlike previous zero-shot dubbing systems that can only generate speech given a reference speech, our method can generate joint speech and sound via text description and optional reference speech. }
  \label{holidubber_overview}
\end{figure}

To address these limitations, we propose HoliDubber, which is illustrated in Figure \ref{holidubber_overview}, a holistic video dubbing framework that jointly generates speech and sound effects from a single text description within a unified generative process. We first construct a scalable audio captioning pipeline that leverages a large audio language model to annotate each training sample with structured descriptions covering speaker profile, speech instruction, and non-speech acoustic content. We then train an Audio-VAE on various audio data, spanning speech, environmental sounds, music, and their mixtures to establish a shared continuous latent space for complex acoustic scene modeling. Built upon this representation, HoliDubber adopts a patch-based autoregressive diffusion architecture inspired by \cite{pmlr-ditar}, where a causal language model captures global temporal structure over aggregated patch embeddings and a Diffusion Transformer decoder generates high-fidelity continuous tokens within each patch. Visual features are integrated via cross-attention to synchronize the generated audio with the speaker's articulation. We further devise a multi-stage training strategy that progresses from large-scale text-to-audio pre-training to structured-prompt fine-tuning, where auxiliary text fields are randomly dropped during training, naturally enabling both zero-shot dubbing with reference speech and text-prompt-guided dubbing at inference time. Extensive experiments demonstrate that HoliDubber significantly outperforms existing methods in speech quality, synchronization, and speaker similarity, while results on our proposed HoliDub-Bench further validate its effectiveness in complex acoustic scenes.

Our major contributions can be summarized as follows:
\begin{itemize}
\item We present HoliDubber, a holistic video dubbing framework that unifies speech and sound effect generation within a single text-guided generative process, moving beyond the speech-only paradigm of existing dubbing systems.
\item We propose a patch-based audio-video fusion mechanism via cross-attention for fine-grained synchronization between generated audio and the speaker's visual cues.
\item We devise a multi-stage training pipeline that progresses from large-scale text-to-audio pre-training to structured-prompt fine-tuning with random text field dropout, enabling a single model to flexibly operate in both zero-shot dubbing and text-prompt-guided dubbing modes.
\item We construct a scalable audio captioning pipeline that produces structured, multi-field textual descriptions for diverse audio content, and curate HoliDub-Bench, a benchmark with fine-grained acoustic annotations for holistic video dubbing evaluation under complex acoustic environments.
\end{itemize}

\section{Related Work}

\subsection{Text-to-Speech Synthesis}
Modern Text-to-Speech (TTS) research has split into two primary paradigms: discrete token-based modeling and continuous representation based modeling.

\textbf{Discrete Token Based TTS.} Influenced by the success of Large Language Models (LLMs), this paradigm treats speech synthesis as a generative sequence task. Early works like AudioLM \cite{borsos2023audiolm} and VALL-E \cite{wang2023valle} utilized neural audio codecs \cite{defossez2022encodec} to convert speech into discrete acoustic tokens via Residual Vector Quantization. To enhance efficiency and scalability, non-autoregressive (NAR) models such as SoundStorm \cite{borsos2023soundstorm} and in-context learners like VoiceCraft \cite{peng2024voicecraft} were introduced. Recent advancements, including CosyVoice \cite{du2024cosyvoice,du2024cosyvoice2} and FireRedTTS \cite{guo2024fireredtts}, have achieved human-parity quality and low-latency streaming by incorporating text-based LLM initialization and large-scale data pre-training. Others, such as IndexTTS 2 \cite{zhou2026indextts2}, have focused on precise duration control. However, discrete approaches often suffer from inherent quantization artifacts, which can limit overall acoustic fidelity.

\textbf{Continuous Representation Based TTS.} To circumvent the information loss in quantization, this paradigm directly models continuous features like mel-spectrograms or audio latents. Following the foundational Tacotron 2 \cite{tacotron2} and FastSpeech \cite{ren2019fastspeech}, recent autoregressive models like MELLE  \cite{meng2025melle} generate continuous frames directly to capture superior prosody. Concurrently, diffusion and flow-matching techniques have been integrated into NAR frameworks \cite{chen2025f5tts,le2023voicebox,guan2024reflowtts, mehta2024matchatts} to enhance detail and diversity. Hybrid architectures, such as ARDiT \cite{liu2024ardit}, DiTAR \cite{pmlr-ditar}, and VibeVoice \cite{peng2025vibevoice}, leverage Diffusion Transformers \cite{dit} (DiT) within an autoregressive framework to unify semantic coherence with acoustic naturalness.

Our framework extends DiTAR by incorporating visual-spatial cues, enabling it to tackle the challenges of automated video dubbing. This cross-modal integration substantially broadens the application potential of the original architecture.

\subsection{Automated video dubbing}

Automated video dubbing aims to synthesize high-quality speech conditioned on both text and visual inputs. Early research primarily focused on adapting TTS models to generate lip-synced audio using lip-cropped video features. Representative methods include NeuralDubber \cite{hu2021neuraldubber}, which utilizes a text-video aligner to bridge cross-modal gap; HPMDubber \cite{cong2023hpmdub}, which introduces a hierarchical duration aligner for phoneme-to-frame mapping; and StyleDubber \cite{cong2024styledubber}, which employs phoneme-guided lip alignment to regulate speech timing. However, these approaches often enforce rigid temporal constraints between speech and video, which can compromise pronunciation accuracy and prosodic naturalness, leading to a large performance gap compared to state-of-the-art TTS systems \cite{du2024cosyvoice, pmlr-ditar}.

In response to these challenges, a more recent paradigm has shifted the focus toward zero-shot dubbing, prioritizing acoustic fidelity, speaker similarity, and emotional expressivity. Initial efforts in this direction, such as Speaker2Dubber \cite{zhang2024speaker2dubber}, EmoDubber \cite{cong2025emodubber}, and ProDubber \cite{zhang2025produbber}, typically build upon FastSpeech-based backbones, enhancing synthesis through large-scale pre-training, flow-based emotion control, or prosody-disentangled adaptation. More recently, advanced generative frameworks have further pushed the boundaries of realism; for instance, FlowDubber \cite{cong2025flowdubber} integrates dual contrastive learning with LLM and flow matching. More recently, the integration of Multimodal Large Language Models (MLLMs) has paved the way for richer cross-modal alignment. Works such as DeepDubber-V1 \cite{zheng2025deepdubber}, FunCineForge \cite{liu2026funcineforge} and InstructDubber \cite{zhang2026instructdubber} employ Chain-of-Thought (CoT) reasoning or natural language instructions to infer nuanced dubbing styles and emotional cues, offering a more interpretable control interface for complex, real-world scenarios.
Despite these advancements, existing models predominantly focus on speech synthesis while neglecting the accompanying environmental acoustics, or they treat speech and sound effects as decoupled components. To address this, we enrich the textual instructions with detailed audio captions. This allows the model to leverage a unified prompting interface to simultaneously synthesize intelligible speech and synchronized sound effects, ensuring a holistic and immersive dubbing experience that aligns with both linguistic content and visual dynamics.

\subsection{Text-to-Audio Generation}
Text-guided audio generation has gained significant traction across diverse applications, including gaming sound effects and virtual reality. Existing Text-to-Audio (TTA) methods are generally categorized into two paradigms: discrete token-based autoregressive generation and diffusion-based continuous modeling.
The first paradigm treats audio synthesis as a sequence prediction task over discrete acoustic tokens \cite{kreuk2022audiogen, agostinelli2023musiclm}. In contrast, the second paradigm leverages the generative prowess of diffusion models to produce high-fidelity mel-spectrograms or audio latent representations \cite{pmlr-audioldm,guan2024lafma,li2025meanaudio}. For instance, DiffSound \cite{yang2023diffsound} employs a non-autoregressive diffusion model over discrete codes derived from a VQ-VAE \cite{van2017vqvae}. AudioLDM \cite{pmlr-audioldm} and its successor, AudioLDM2 \cite{liu2024audioldm2}, utilize pretrained contrastive text-audio representations (e.g., CLAP \cite{elizalde2023clap}) and AudioMAE \cite{huang2022audiomae} features to condition a Latent Diffusion Model (LDMs), effectively reducing the reliance on large-scale paired data.
To further enhance semantic understanding, Tango \cite{ghosal2023tango} incorporates instruction-tuned LLMs, Flan-T5 \cite{chung2024flant5}, to capture intricate cross-modal correlations, challenging the necessity of a shared latent space. Auffusion \cite{xue2024auffusion} 
adapts powerful pretrained Text-to-Image LDMs to inherit robust generative priors for audio synthesis. 
Existing TTA models often neglect intelligible speech. We enhance DiTAR \cite{pmlr-ditar} by expanding its input descriptions, empowering the model to generate joint speech and sound effects.

\subsection{Video-conditioned Soundtrack Generation}

Recent advancements have sought to integrate video-conditioned speech and sound synthesis into unified frameworks. Industrial models such as Meta’s Audiobox \cite{vyas2023audiobox} and Google’s Video-to-Audio (V2A) leverage diffusion backbones to generate audio from multimodal prompts, including video, text, and speech clues. 
Despite these strides, unified generation in the academic community remains bottlenecked by data scarcity and architectural complexity. For instance, AudioGenOmni \cite{wang2025audiogenomni} utilizes in-context conditioning within a flow-based model but fails to achieve simultaneous speech and sound generation. While DeepAudio \cite{zhang2025deepaudio} and DualDub \cite{tian2025dualdub} integrate specialized heads into LLM backbones, they often rely on intricate multi-stage curriculum learning to mitigate performance degradation during joint training. This reliance on fragmented pipelines leaves the potential for efficient, end-to-end joint generation largely underexplored.
While direct V2A captures temporal synchronization, it often lacks the flexibility required for diverse creative scenarios. 
Consequently, a purely vision-conditioned model is incapable of generating auxiliary sound effects that are not visually present. To overcome this limitation, Our HoliDubber advocates for a text-prompt-guided approach. By integrating natural language descriptions, our framework allows users to flexibly inject specific acoustic effects into a visually clean scene, significantly enhancing the controllability and creative potential of video dubbing.

\begin{figure*}[t]
  \centering
  \vspace{-12pt}
  \includegraphics[width=0.86\linewidth]{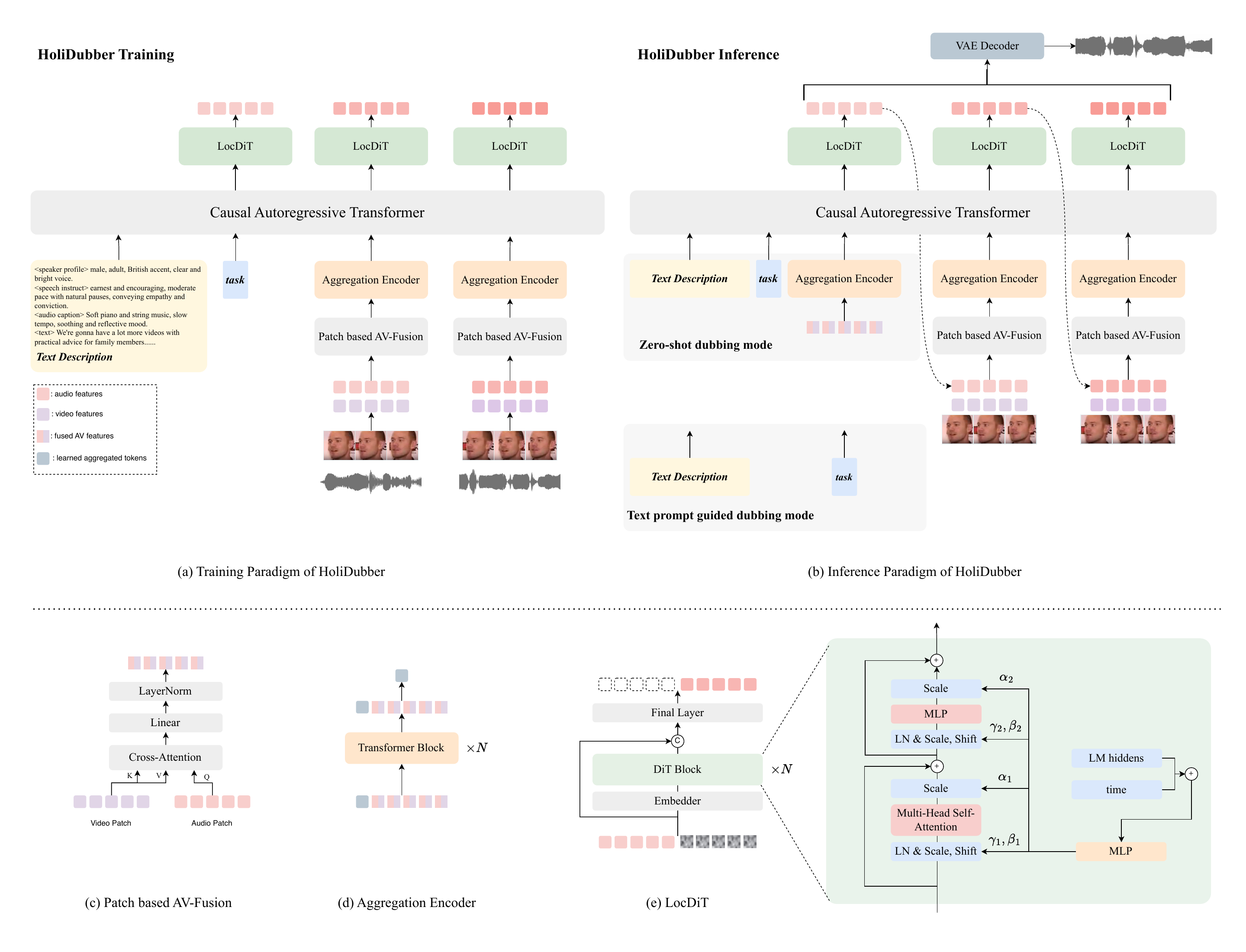}
  \caption{Overall framework of HoliDubber. (a) illustrates the training paradigm of HoliDubber. (b) illustrates the inference paradigm of HoliDubber. (c), (d), (e) depict the overall architecture of patch based AV-fusion module, aggregation encoder and local diffusion transformer (LocDiT), respectively.}
  \label{holidubber_arc}
\end{figure*}

\section{HoliDubber}

\subsection{Framework Overview}

HoliDubber aims to simultaneously synthesize complex acoustic scenes $Y$ that encompass both speech and environmental sound effects. It supports two versatile inference modes: (1) zero-shot video dubbing guided by a reference speech \(R\) to preserve speaker identity , and (2) text-prompt-guided generation, which synthesizes synchronized acoustic content solely based on natural language descriptions \(T\).
Both inference modes require a text prompt \(T\) and a silent video \(V\) as inputs. We can formulate the task as:
\begin{equation}
    Y = HoliDubber([R], T, V)
\end{equation}
where [R] indicates that the reference speech is optional.
As shown in Figure \ref{holidubber_arc} (b), HoliDubber consists of five components. Specifically, the VAE is employed to encode audio into latent features, acting as the primary processing units for the subsequent generative model. These representations are finally mapped back to the acoustic space via the VAE decoder.  The patch based Audio-Visual Fusion Module merges localized video patch features with corresponding audio features to produce fused audio-visual representations.  
By prepending a learnable token to the fused audio-visual sequence, the Aggregation Encoder captures the holistic multi-modal context and produces a compressed representation at the learnable token's position.
By conditioning on the learnable compressed representations and textual inputs, the Causal Autoregressive Transformer produces an in-context latent space to guide subsequent synthesis.  
Utilizing a flow-matching objective, the Local Diffusion Transformer (LocDiT) synthesizes the audio representation of the current patch by leveraging the in-context embeddings from the Causal Autoregressive Transformer and the previously generated patch as joint conditioning signals.

\subsection{Patch based Audio-Visual Fusion}
\label{sec_av_fusion}
The patch-based Audio-Visual Fusion Module is designed to establish fine-grained correspondences between audio signals and visual articulation dynamics at the patch level. As shown in Figure \ref{holidubber_arc} (c), we fuse patch-level visual features from the video encoder with audio features via cross-attention layers, where audio features serve as queries and video patch features serve as keys and values. Notably, each audio patch attends to the visual features of its forthcoming temporal segment rather than the current one, allowing the model to anticipate the speaker's upcoming articulation and produce temporally aligned audio in an autoregressive generation setting.
We adopt cross-attention over alternative fusion strategies such as feature concatenation for two reasons. First, cross-attention acts as a side-channel injection that modulates audio representations through selective visual queries without altering the dimensionality or distributional properties of the audio feature sequence, thereby preserving the generative capabilities inherited from the pre-trained text-to-audio backbone. Second, the attention mechanism enables each audio token to adaptively focus on the most informative visual cues, such as key articulatory moments, rather than uniformly incorporating all visual features. This selective attention is critical for capturing the sparse and temporally varying nature of audio-visual correspondences in real-world dubbing scenarios, where only a subset of video frames carry significant lip-movement information at any given moment.

\subsection{Aggregation Encoder}

After the patch-based fusion stage, the resulting sequence contains multiple fused audio-visual tokens encoding localized cross-modal interactions. While this distributed representation preserves fine-grained detail, downstream tasks typically require a single compact vector summarizing the entire multi-modal context. To this end, we adopt the Aggregation Encoder from DiTAR \cite{pmlr-ditar}, which prepends a learnable special token to the beginning of the input sequence and feeds it through a bidirectional Transformer encoder, where the output at the special token's position serves as the aggregation embedding \cite{devlin2019bert}. As shown in Figure \ref{holidubber_arc} (d), this mechanism allows the special token to attend to all fused audio-visual positions and progressively distill the most salient cross-modal cues into a fixed-dimensional representation in a content-adaptive manner, effectively compressing the patch-level multi-modal sequence into a holistic embedding suitable for downstream processing.

\subsection{Local Diffusion Transformer}
Following the divide-and-conquer strategy proposed in DiTAR \cite{pmlr-ditar}, our framework delegates inter-patch and intra-patch modeling to two complementary modules. While the Causal Autoregressive Transformer captures long-range dependencies across patches, it only produces a single coarse embedding per patch, which is insufficient to recover the fine-grained, high-dimensional structure within each local patch of audio tokens. As shown in Figure \ref{holidubber_arc} (e), the Local Diffusion Transformer (LocDiT) addresses this by employing a bidirectional Transformer with a flow-matching objective to synthesize the full continuous representation of the current patch. Unlike the unidirectional causal attention used in the autoregressive backbone, LocDiT leverages bidirectional attention to model the strong inter-frame correlations that exist among adjacent continuous tokens within a local region. This yields an inherent coarse-to-fine generation process: the autoregressive model first compresses each patch into an implicit coarse feature, and LocDiT subsequently expands it into high-fidelity continuous tokens in an end-to-end manner, avoiding the cumulative errors typical of explicit multi-stage coarse-to-fine pipelines. To further improve generation quality, historical audio patches of tokens are utilized as prefix inputs for the LocDiT,  providing richer contextual continuity at patch boundaries. Concretely, the output embedding from the autoregressive model and the previously generated patch jointly condition LocDiT, enabling it to produce audio representations that are both locally coherent and globally consistent with the multi-modal context.

\subsection{Training Objective}

As shown in Figure \ref{holidubber_arc} (a), given paired audio-visual training samples, we first extract continuous audio representations from the audio encoder and patch-level visual features from the video encoder. The audio and video representation are segmented into non-overlapping patches of size $P$. For each audio patch, the corresponding video patch features are fused via the Patch-based Audio-Visual Fusion Module, which employs cross-attention layers with audio tokens as queries and video patches as keys and values, producing a sequence of fused audio-visual tokens that capture fine-grained cross-modal correspondences. The fused tokens within each patch, together with a prepended learnable aggregation token, are then processed by the Aggregation Encoder, whose output at the aggregation token's position yields a single embedding that summarizes the local multi-modal context of that patch.

The resulting sequence of aggregation embeddings is fed into the Causal Autoregressive Transformer, which models inter-patch dependencies via next-token prediction with causal attention. At each patch position $i$, the autoregressive model produces an output embedding $h_i$ that encodes all preceding multi-modal context $(x_1, x_2, \ldots, x_i)$. This embedding, along with the previously generated audio patch, serves as the conditioning signal for the Local Diffusion Transformer (LocDiT), which is responsible for generating the fine-grained audio tokens within the next patch.

We train the LocDiT with a conditional flow matching \cite{lipman2022flow} objective. Specifically, we construct a linear interpolation between a noise sample $\epsilon \sim \mathcal{N}(0, \mathbf{I})$ and the ground-truth audio tokens $x_0$ of the target patch as $x_t = (1-t)\epsilon + tx_0$, where $t \sim \mathcal{U}(0,1)$. The LocDiT is optimized to predict the velocity field $v_\theta(x_t, t, c)$ that transports the noise distribution toward the data distribution, where $c = (h_i, x_{i-1})$ denotes the joint condition. The overall training loss is:

\begin{equation}
\mathcal{L} = \mathbb{E}_{t, \epsilon, x_0}\left[\|v_\theta(x_t, t, c) - (x_0 - \epsilon)\|^2\right]
\end{equation}

The entire model, including the Patch based Audio-Visual Fusion module, Aggregation encoder, Causal Autoregressive Transformer, and LocDiT, is trained end-to-end by minimizing this single flow matching objective. At inference time, we generate each audio patch autoregressively, the causal model produces the conditioning embedding for the current step, and LocDiT synthesizes the corresponding patch by sampling $\epsilon \sim \mathcal{N}(0, \mathbf{I})$ and solving the learned ODE from $t=0$ to $t=1$ using an off-the-shelf numerical solver.

\section{Experimental Setup}
\subsection{Datasets}
We train our model on the VoxCeleb 2 \cite{chung2018voxceleb2} dataset and  CelebV-Dub \cite{sung2025voicecraft} dataset. 

\textbf{VoxCeleb 2} VoxCeleb2 dataset is originally a large-scale audio-visual speaker recognition corpus collected from YouTube videos. VoxCeleb2 contains over 1 million utterances from 6112 speakers, spanning approximately 2442 hours of speech.  Crucially, all segments are captured in the wild, with videos shot in a large number of challenging visual and auditory environments, including red carpet interviews, outdoor stadiums, indoor studios, speeches given to large audiences, and crude videos shot on hand-held devices.  The visual modality similarly exhibits unconstrained conditions with variations in pose, lighting, image quality, and motion blur. These characteristics make VoxCeleb2 particularly suitable for our task, as the diverse acoustic environments and natural background noise require the model to learn robust audio-visual correspondences under realistic conditions.  

For the VoxCeleb2 dataset, we reserve a total of 991 clips for evaluation: 641 clips are selected to construct HoliDub-Bench (detailed in Sec. \ref{bench}), and an additional 350 clips are randomly sampled as a standard test set for general evaluation. The remaining clips are used for training.

\textbf{CelebV-Dub.} CelebV-Dub \cite{sung2025voicecraft} is a curated dataset of expressive, real-world videos specifically designed for automated video dubbing, comprising approximately 86 hours of speech. The dataset features unconstrained in-the-wild settings with rich emotional variations, diverse speaking styles, and challenging acoustic conditions, making it particularly suitable for evaluating dubbing models that must generate speech aligned with expressive facial movements. 
For the CelebV-Dub dataset, we reserve a total of 477 clips for evaluation: 359 clips are selected to construct HoliDub-Bench (detailed in Sec. \ref{bench}), and an additional 118 clips are randomly sampled as a standard test set for general evaluation. The remaining clips are used for training.

% \vspace{-3pt}
\subsection{Audio Caption Pipeline}
\label{pipeline}

We build an automated audio caption pipeline to generate structured, multi-dimensional annotations for each audio clip. The pipeline leverages Qwen3-Omni-30B-A3B-Instruct~\cite{xu2025qwen3omni} as the backbone captioning model and Whisper-large-v3 as the speech recognition (ASR) model.
% \vspace{-3pt}

\textbf{Structured Caption Format.}
For each audio clip, the Qwen3-Omni model generates a structured caption comprising three fields:

\texttt{<speaker profile>}: speaker attributes including gender, age range, accent, and vocal timbre, indexed by speaker identity;

\texttt{<speech instruct>}: prosodic description covering speaking rate, emotional tone, and vocal qualities such as breathiness or trembling;

\texttt{<audio caption>}: non-speech audio analysis describing background music (instruments, melody, atmosphere), ambient sounds (physical sound sources and characteristics), and spatial dynamics (loudness variation, distance);

Additionally, we obtain the \texttt{<text>} field through a separate pipeline: source separation \cite{wang2023melroformer} is first applied to isolate the speech component, followed by Whisper \cite{radford2023whisper} based automatic speech recognition (ASR) to generate the transcription.

\textbf{Prompt Design.}
We design a captioning prompt that instructs the model to describe the speech content and the background sound/music present in the audio. The prompt enforces a strict "what you hear is what you describe" principle, explicitly prohibiting the model from hallucinating non-existent sound elements, ensuring that the generated captions faithfully reflect the actual acoustic content of each clip.
The details of the audio caption prompt design is described in Appendix B.

\subsection{HoliDub-Bench}
\label{bench}

To provide a more challenging and comprehensive evaluation, we construct HoliDub-Bench, a curated benchmark comprising 1,000 clips (approximately 2.4 hours) drawn from two sources. From VoxCeleb2, we select samples containing rich background sounds such as music, ambient noise, and overlapping audio. From CelebV-Dub, we select samples exhibiting pronounced emotional variations, including dramatic shifts in tone, intensity, and expressiveness. All selected clips are processed through the structured captioning pipeline described in Sec \ref{pipeline}, yielding text-video-audio triplets. Since these clips are specifically curated for complex acoustic scenes encompassing both speech and non-speech elements, we evaluate exclusively under the text-prompt-guided dubbing mode on HoliDub-Bench, as only this mode conditions generation on the full structured description including the \texttt{<audio caption>} field required to reproduce the target acoustic environment. The details of HoliDub-Bench are described in Appendix A.

\subsection{Implementation Details}
We first train an Audio-VAE to encode raw audio waveforms into continuous latent representations. Existing open-source audio VAEs \cite{evans2024stableaudio, niu2025semanticvae} are typically trained exclusively on either general audio data or speech data, limiting their ability to simultaneously model both speech and non-speech acoustic content. To overcome this limitation, we train a unified Audio-VAE on heterogeneous audio data spanning speech, environmental sounds, music, and their natural mixtures. Architecturally, our Audio-VAE largely follows the backbone design of DAC \cite{kumar2023dac}, incorporating ResidualUnits, Snake activations, weight-normalized convolutions, and multi-scale dilations.
The key distinction lies in the bottleneck design: while DAC employs Residual Vector Quantization to produce discrete tokens, our Audio-VAE replaces it with a VAE-style bottleneck parameterized by mean and log-variance, yielding continuous latent representations that are more naturally suited for flow matching  generative modeling. 
During Audio-VAE training, all audio is resampled to 32 kHz, and the encoder compresses the input into continuous latents at a frame rate of 25 Hz. The training data comprises approximately 30,000 hours in total, evenly distributed across three categories in a 1:1:1 ratio: speech, general audio (including environmental sounds and music), and naturally mixed data where speech co-occurs with background sounds in the original recordings.

Next, we trained a text-guided audio generation model without video input, following the DiTAR \cite{pmlr-ditar} framework. The Causal Autoregressive Transformer is initialized from a pre-trained Qwen2.5-1.5B language model \cite{yang2024qwen25}. The training data totals approximately 130,000 hours, comprising: 80,000 hours of speech from the Emilia \cite{he2024emilia} dataset with plain text transcriptions; 20,000 hours of naturally mixed speech-audio data; 20,000 hours of speech annotated with <speaker profile> and <speech instruct> attributes; and 10,000 hours of non-speech audio and music. With the exception of Emilia, all other audio data are annotated using the structured captioning pipeline described in Sec \ref{pipeline}.

Finally, we train HoliDubber on top of the pre-trained text-guided audio generation model. All input videos are processed at 25 FPS, and we employ AV-HuBERT \cite{shi2022avhubert} as the video encoder to extract visual features, naturally aligned with the 25 Hz frame rate of the audio latents produced by the Audio-VAE. The patch size is set to 5. The Aggregation Encoder consists of 4 Transformer blocks with 8 attention heads, and the LocDiT comprises 12 Transformer blocks with 16 attention heads. For both VoxCeleb2 and CelebV-Dub, all audio is resampled to 32 kHz, and the training splits of both datasets are combined for joint training.   During training, we apply the selective text-prompt dropout strategy with a probability of 50\%, where the auxiliary fields (\texttt{<speaker profile>}, \texttt{<speech instruct>}, \texttt{<audio caption>}) are dropped and only the \texttt{<text>} transcription is retained, enabling the model to support both zero-shot and text-prompt-guided inference modes.
The inference step of  LocDiT is set to 10.

\vspace{-3pt}
\subsection{Evaluation Metrics}
The evaluation metrics vary depending on the inference mode.

\textbf{Zero-shot Dubbing mode.}
For the zero-shot dubbing mode, the primary objective is to generate speech that preserves the vocal timbre of the reference speaker while ensuring naturalness and high quality. We evaluate performance across multiple dimensions: LSE-C/D \cite{chung2016syncnet} for lip synchronization accuracy, SPK-SIM for speaker similarity between the generated and reference speech following \cite{chen2025f5tts}, UTMOS \cite{saeki2022utmos} for overall speech quality, Word Error Rate (WER) for content accuracy obtained by transcribing the synthesized speech with Whisper-large-v3, and Emo-SIM for emotion similarity computed as the cosine similarity between emotion embeddings extracted by Emotion2Vec \cite{ma2024emotion2vec} from the generated and reference audio pairs. In addition, we conduct Mean Opinion Score (MOS) tests for subjective human evaluation.

\begin{table*}[t]
\centering
\footnotesize
\caption{Evaluation of generated audio in the zero-shot dubbing mode on the VoxCeleb 2 and CelebV-Dub test sets.}
\label{table_zeroshot_mode}
\vspace{-6pt}
\resizebox{0.95\linewidth}{!}
{
\begin{tabular}{lcccccccccccccc}
\toprule
\multirow{2}{*}{Model} 
& \multicolumn{7}{c}{VoxCeleb 2 test set} 
& \multicolumn{7}{c}{CelebV-Dub test set} \\
\cmidrule(lr){2-8} \cmidrule(lr){9-15}
& LSE-C$\uparrow$ & LSE-D$\downarrow$ & SPK-SIM$\uparrow$ & EMO-SIM$\uparrow$ & WER$\downarrow$ & UTMOS$\uparrow$ & MOS$\uparrow$ 
& LSE-C$\uparrow$ & LSE-D$\downarrow$ & SPK-SIM$\uparrow$ & EMO-SIM$\uparrow$ & WER$\downarrow$ & UTMOS$\uparrow$ & MOS$\uparrow$  \\
\midrule
GT      & - & - & \textbf{0.71} & 0.980  & \textbf{19.91}  & 2.51 & 3.82 & - & - & \textbf{0.55} & 0.950  & \textbf{9.19}  & 2.87 &  3.92 \\
\midrule
AlignDiT \cite{choi2025aligndit}     & \textbf{6.92} & \underline{7.90} & 0.55 & \underline{0.980}  & \underline{20.92}  & 2.77 & 3.72 & \underline{6.63} & \textbf{7.66} & 0.41 & 0.951  & \underline{12.31}  & 3.16 &  3.79 \\
VoiceCraft-Dub \cite{sung2025voicecraft}    & 4.50 & 9.95 & 0.31 & 0.973  & 49.12  & 2.64 & 3.62 & 6.31 & 8.33 & 0.27 & \textbf{0.955}  & 29.03  & \underline{3.27} &  3.71 \\
FunCineForge \cite{liu2026funcineforge}   & 1.56 & 13.29 & 0.46 & 0.978  & 21.22  & \textbf{3.70} & \textbf{3.91} & 2.63 & 11.89 & 0.34 & \underline{0.954}  & 21.47  & \textbf{3.65} &  \textbf{4.06} \\
HoliDubber    & \underline{6.83} & \textbf{7.87} & \underline{0.68} & \textbf{0.981}  & 24.12  & \underline{2.79} & \underline{3.83} & \textbf{6.65} & \underline{7.92} & \underline{0.46} & 0.951  & 17.95  & 3.24 &  \underline{3.96} \\
\bottomrule
\end{tabular}
}
\end{table*}

\begin{table*}[h]
\centering
\footnotesize
\caption{Evaluation of generated audio in the text prompt guided dubbing mode on the VoxCeleb2 and CelebV-Dub test sets. }

\label{table_prompt_mode}
\vspace{-9pt}
% \resizebox{1.0\linewidth}{!}
% {
\begin{tabular}{llccccccccccc}
\toprule
Dataset & Model & LSE-C$\uparrow$ & LSE-D$\downarrow$ & EMO-SIM$\uparrow$ & WER$\downarrow$ & UTMOS$\uparrow$ & MOS$\uparrow$ & FD$\downarrow$ & FAD$\downarrow$ & KLD$\downarrow$ & IS$\uparrow$ \\
\midrule
\multirow{3}{*}{VoxCeleb2}
& GT & - & - & - & \textbf{19.91} & 2.51 & 3.82 & - & - & - & - \\
& FunCineForge & 1.54 & 13.17 & 0.990 & 21.10 & \textbf{3.55} & 3.86 & 16.34 & 5.18 & 1.60 & 1.16 \\
& HoliDubber   & \textbf{6.52} & \textbf{8.18} & \textbf{0.994} & 20.03 & 3.04 & \textbf{3.92} & \textbf{10.21} & \textbf{3.82} & \textbf{1.43} & \textbf{1.26} \\
\midrule
\multirow{3}{*}{CelebV-Dub}
& GT & - & - & - & \textbf{9.19} & 2.87 & 3.92 & - & - & - & - \\
& FunCineForge & 2.69 & 11.84 & 0.977 & 22.17 & \textbf{3.65} & 3.96 & 15.24 & 4.11 & 0.79 & 1.15 \\
& HoliDubber   & \textbf{6.55} & \textbf{8.06} & \textbf{0.978} & 19.42  & 3.12 &  \textbf{3.99} & \textbf{6.85} & \textbf{3.16} & \textbf{0.69} & \textbf{1.32} \\
\bottomrule
\end{tabular}
% }
\end{table*}

\textbf{Text Prompt guided Dubbing mode.}
For the text prompt-guided dubbing mode, the only difference is that no reference speech is available for comparison; instead, we use the ground-truth audio as the reference for EMO-SIM. Therefore, all metrics described above except SPK-SIM remain applicable. Additionally, since the text descriptions contain audio caption-related content describing background sounds and music, we incorporate Fréchet Distance (FD), Fréchet Audio Distance (FAD), Kullback-Leibler Divergence (KLD) and Inception Score (IS) to evaluate the quality of the generated non-speech audio, following AudioLDM \cite{pmlr-audioldm}.

\subsection{Baseline Models}

We compare our method with three open-source approaches: AlignDiT \cite{choi2025aligndit}, VoiceCraft-Dub \cite{sung2025voicecraft}, and FunCineForge \cite{liu2026funcineforge}, all of which are zero-shot video dubbing models trained on relatively large-scale datasets. Among these baselines, FunCineForge also supports a text-prompt-guided dubbing mode; however, unlike our framework, it still requires a reference speech as a mandatory input even in this mode, limiting its flexibility in scenarios where no reference speaker is available.

\textbf{AlignDiT \cite{choi2025aligndit}} AlignDiT is a DiT \cite{dit} based approach for automatic video dubbing, which extends non-autoregressive zero-shot TTS models \cite{chen2025f5tts} to incorporate visual cues as inputs. It was trained on the LRS3 \cite{afouras2018lrs3} dataset.

\textbf{VoiceCraft-Dub \cite{sung2025voicecraft}.} VoiceCraft-Dub is a Neural Codec Language Model (NCLM) based approach for automatic video dubbing, which extends the high-quality, human-like speech synthesis capabilities of NCLMs to incorporate visual cues as inputs. It was trained on the LRS3 \cite{afouras2018lrs3} dataset and CelebV-Dub dataset.

\textbf{FunCineForge \cite{liu2026funcineforge}.} FunCineForge is an MLLM-based dubbing model designed for diverse cinematic scenes, which integrates multimodal information including facial features, text, and audio references with a flow matching module supporting flexible speaker switching. It was primarily trained on CineDub-CN and CineDub-EN, a large-scale television dubbing dataset constructed through its own end-to-end data pipeline with rich annotations.

In addition to the above baselines, we include our pre-trained text-guided audio generation model (denoted as \textbf{TTA}) for comparison on HoliDub-Bench. This comparison isolates the contribution of visual information in our framework.

Note that earlier dubbing methods such as StyleDubber \cite{cong2024styledubber}, ProDubber \cite{zhang2025produbber} and InstructDubber \cite{zhang2026instructdubber},  are trained on small single-dataset benchmarks such as Grid \cite{cooke2006grid}, and evaluating them on our test sets would be unfair due to the significant domain gap.

\section{Experimental Results}

\begin{table*}[!tbp]
\centering
\footnotesize
\caption{Evaluation of generated audio  on  HoliDub-Bench. }
\label{table_holibench}
\vspace{-6pt}
\resizebox{0.66\linewidth}{!}
{
\begin{tabular}{lccccccccccc}
\toprule
& Model & LSE-C$\uparrow$ & LSE-D$\downarrow$ & EMO-SIM$\uparrow$ & WER$\downarrow$ & UTMOS$\uparrow$ & MOS$\uparrow$ & FD$\downarrow$ & FAD$\downarrow$ & KLD$\downarrow$ & IS$\uparrow$ \\
\midrule
& GT & - & - & - & 14.51 & 2.80 & 3.89 & - & - & - & - \\
& TTA & 1.21 & 15.26 & \textbf{0.993} & 15.28 & 2.97 & 3.81 & \textbf{9.52} & 10.51 & 1.89 & \textbf{1.54} \\
& HoliDubber   & \textbf{6.44} & \textbf{8.08} & 0.992 & \textbf{12.81} & \textbf{3.02} & \textbf{3.96} & 10.95 & \textbf{3.08} & \textbf{1.86} & 1.51 \\
\bottomrule
\end{tabular}
}
\end{table*}

\subsection{Zero-shot Video Dubbing Mode Results}
\label{exp_zero-shot}
As shown in Table \ref{table_zeroshot_mode}, we analyze the zero-shot video dubbing results from five aspects.

\textbf{Lip synchronization.} HoliDubber and AlignDiT achieve the strongest lip-sync performance with closely matched scores on both datasets. FunCineForge exhibits substantially lower accuracy due to its sensitivity to timestamp annotations.

\textbf{Speaker similarity.} HoliDubber attains the highest SPK-SIM among all methods, most closely approaching the GT upper bound, indicating that our continuous representation and patch based fusion effectively preserve speaker identity. VoiceCraft-Dub scores lowest, suggesting limited generalization of its NCLM framework.

\textbf{Emotion similarity.} All methods maintain EMO-SIM above 0.95. HoliDubber achieves the best or competitive scores on both datasets, demonstrating that our model captures expressive nuances from visual cues.

\textbf{Content accuracy.} AlignDiT yields the lowest WER, closely tracking the GT. HoliDubber exhibits moderately higher WER, which we attribute to our holistic generation paradigm: the co-occurring non-speech acoustic elements (e.g., ambient noise, music) in the jointly synthesized audio stream can interfere with Whisper-based ASR transcription, inflating the measured WER beyond what the actual speech intelligibility would suggest. 

\textbf{Speech quality.} FunCineForge achieves the highest UTMOS and MOS, substantially exceeding the GT. However, UTMOS closer to the GT is more desirable than simply higher scores, as the GT reflects natural acoustic characteristics of real-world recordings. An UTMOS significantly exceeding the GT suggests over-smoothed, studio-clean speech that deviates from the acoustic realism of the original scene. We further observe a trade-off between speech quality and lip-sync accuracy: without tight frame-level alignment constraints, FunCineForge produces smoother speech at the expense of synchronization. HoliDubber achieves UTMOS closest to the GT while ranking second in MOS, reflecting a better balance between perceptual quality, acoustic realism, and audio-visual alignment.

Overall, HoliDubber achieves the most balanced performance across all dimensions, validating the effectiveness of our patch-based cross-modal fusion in jointly achieving precise audio-visual alignment and high-fidelity timbre reproduction.

\subsection{Text Prompt guided Video Dubbing Mode Results}
As shown in Table \ref{table_prompt_mode}, we evaluate the text prompt guided dubbing mode, where speech is synthesized from textual descriptions rather than a reference utterance. Since no reference speech is available, we replace SPK-SIM with audio generation metrics (FD, FAD, KLD, IS) to assess the quality of generated non-speech audio content. Only FunCineForge is compared, as AlignDiT and VoiceCraft-Dub do not support this mode.

\textbf{Lip synchronization.} Consistent with the zero-shot results, HoliDubber significantly outperforms FunCineForge on both datasets, further confirming that the advantage of our patch-based cross-modal fusion is robust across inference modes.

\textbf{Emotion similarity and content accuracy.} HoliDubber achieves higher EMO-SIM on both datasets, indicating better alignment with the emotional characteristics conveyed by the text prompts. HoliDubber also produces lower WER on both sets, demonstrating reliable speech intelligibility under text-guided generation.

\textbf{Speech quality.} FunCineForge achieves higher UTMOS scores, consistent with the quality-alignment trade-off discussed in Sec \ref{exp_zero-shot}. Nevertheless, HoliDubber obtains higher subjective MOS on both datasets, suggesting that human listeners prefer the overall naturalness and expressiveness of our generated speech.

\textbf{Audio generation quality.} HoliDubber substantially outperforms FunCineForge across all audio generation metrics on both datasets, with particularly pronounced improvements in FD and IS. These results demonstrate that HoliDubber generates non-speech audio content—including background sounds and music—significantly closer to the real distribution, benefiting from the holistic audio modeling capability inherited from our pre-trained text-guided audio generation model.

Overall, HoliDubber achieves superior performance across all evaluated dimensions in the text prompt guided mode, demonstrating that our model effectively leverages textual descriptions to synthesize high-quality dubbed audio with both accurate speech content and faithful non-speech acoustic elements.

\begin{table}[!tbp]
  \centering
  \caption{Ablation study of the proposed method in the zero-shot dubbing mode on the VoxCeleb 2  test sets.} 
  \vspace{-6pt}
  \resizebox{1.0\linewidth}{!}
  {
    \begin{tabular}{lcccccccc}
    \toprule
    \# & Methods   & LSE-C $\uparrow$ & LSE-D $\downarrow$ & SPK-SIM$\uparrow$ & EMO-SIM$\uparrow$ & WER $\downarrow$ & UTMOS $\uparrow$  \\
    \midrule
    1 & HoliDubber   & \textbf{6.83} & \textbf{7.87} & \textbf{0.68} & \textbf{0.981} & \underline{24.12} & \textbf{2.79}\\
    \midrule
    2 & w/o ref-video & 4.29 & 10.21 & \underline{0.65} & \underline{0.980} & \textbf{21.43} & \textbf{2.91} \\
    3 & w/o patch-av-fusion & 4.15 & 10.15 & 0.15 & 0.972 & 62.19 & 1.76 \\
    4 & w/o prompt random drop & \underline{4.61} & \underline{9.96} & 0.56 & 0.980 & 25.62 & \underline{2.86} \\
    
    \bottomrule
    \end{tabular}%
    } 
  \label{tab_ablation}%
\end{table}
\vspace{-6pt}

\subsection{Results on HoliDub-Bench}
As shown in Table \ref{table_holibench}, we compare HoliDubber against its pre-trained text-guided audio generation backbone (TTA), which receives identical textual inputs but lacks video conditioning, to directly reveal the contribution of visual information.

HoliDubber achieves dramatically higher lip-sync accuracy than TTA, confirming that our patch-based audio-visual fusion effectively leverages video cues for precise temporal alignment. HoliDubber also attains lower WER than both TTA and GT (12.81 vs 15.28 vs 14.51) along with the highest UTMOS and MOS, suggesting that visual conditioning provides complementary cues that benefit both intelligibility and perceptual quality. For audio generation, TTA achieves slightly better FD and IS, while HoliDubber obtains substantially lower FAD (3.08 vs 10.51), indicating higher fidelity in non-speech audio content. These results validate our design of augmenting a strong text-guided audio backbone with visual information through patch-based cross-modal fusion.

\subsection{Ablation Studies}
As shown in Table \ref{tab_ablation}, we conduct ablations on the VoxCeleb2 test set to isolate the contribution of each key component.

\textbf{w/o ref-video.} Removing video input at inference causes a  lip-sync drop, while WER and UTMOS slightly improve, reflecting the quality–alignment trade-off: without visual constraints, the model gains greater freedom to optimize speech quality at the expense of synchronization.

\textbf{w/o patch-av-fusion.} Replacing cross-attention with feature concatenation~\cite{chen2025f5tts} causes catastrophic degradation, as concatenation alters the input distribution inherited from the pre-trained TTA backbone, confirming that cross-attention is essential for both lip synchronization and overall generation quality.

\textbf{w/o prompt random drop.} Training exclusively with full structured descriptions creates a train-test mismatch in zero-shot mode, where only \texttt{<text>} is provided. This leads to moderate lip-sync degradation. Stochastic field dropout mitigates this by exposing the model to partial inputs during training, enabling robust generalization across both inference modes.

We further compare HoliDubber against a decoupled pipeline alternative in Appendix C, demonstrating the advantage of unified end-to-end speech-and-sound synthesis.

\section{Conclusion}

We introduced HoliDubber, the first holistic framework to unify speech and sound effects generation within a single text-guided gen-
erative process for video dubbing. Utilizing a patch-level cross-attention fusion mechanism and a patch-based autoregressive diffusion transformer architecture, our model effectively grounds audio synthesis in visual articulation dynamics, ensuring superior lip-sync and acoustic realism. 
We further introduced HoliDub-Bench, a curated benchmark specifically designed for evaluating holistic dubbing under challenging acoustic conditions. Extensive experiments demonstrated that HoliDubber significantly outperforms existing methods across multiple evaluation dimensions, establishing a new paradigm for video dubbing in complex acoustic scenes. 
In future work, we plan to extend HoliDubber to support multilingual dubbing, multi-speaker dialogue scenes, and more diverse acoustic scenarios.

% \section{Limitations}

\bibliographystyle{ACM-Reference-Format}
\bibliography{sample-base}

%%
% If your work has an appendix, this is the place to put it.
\clearpage
\appendix

\section{HoliDub-Bench }
\label{app_bench}
\begin{figure}[htbp]
  \centering
  \includegraphics[width=\linewidth]{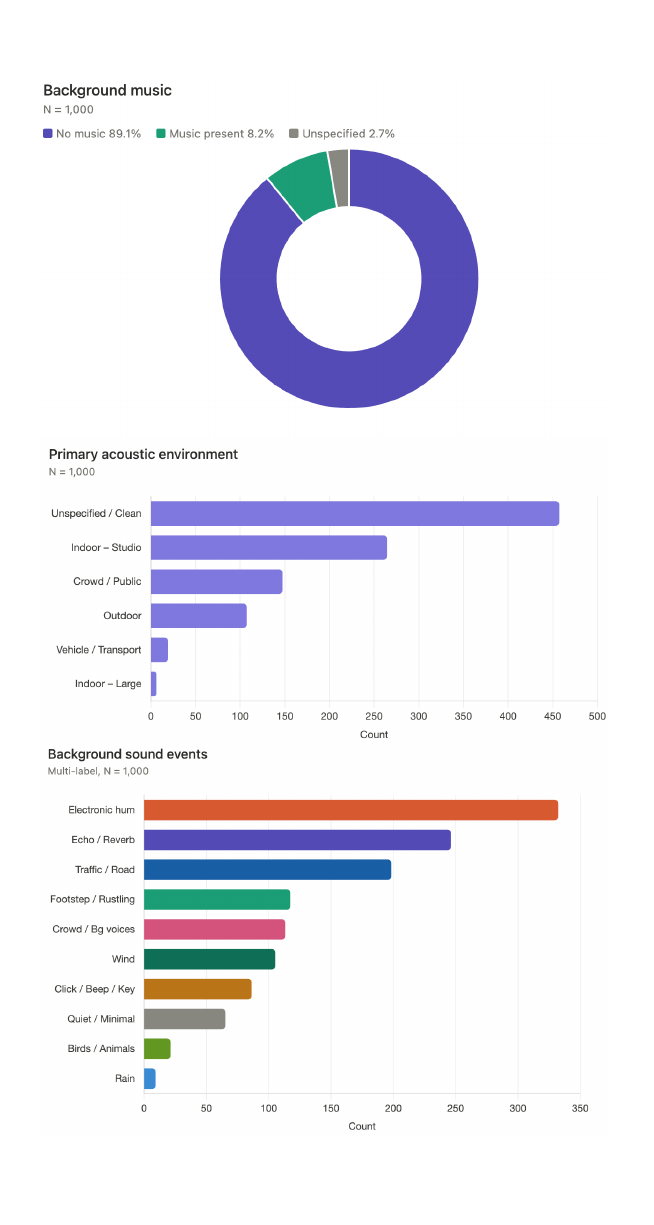}
  \caption{HoliDub-Bench Dataset Statistics.}
  \label{app_bench_sta}
\end{figure}

We provide a detailed characterization of the acoustic conditions in HoliDub-Bench across three complementary dimensions: background music presence, primary acoustic environment, and background sound events.

As shown in Figure \ref{app_bench_sta}, the vast majority of samples (89.1\%) contain no background music, while 8.2\% feature music and 2.7\% remain unspecified. This distribution faithfully mirrors real-world dubbing pipelines, where background music is typically separated or attenuated to isolate dialogue. The inclusion of a meaningful minority of music-present samples further enables evaluation of model robustness under acoustically complex conditions.

Regarding the primary acoustic environment, "Unspecified / Clean" accounts for the largest proportion (45.7\%), followed by "Indoor – Studio/Room" (26.4\%), "Crowd / Public Space" (14.7\%), and "Outdoor" (10.7\%). Less common yet practically important settings such as "Vehicle / Transport" (1.9\%) and "Indoor – Large/Reverberant" (0.6\%) are also covered. This broad environmental coverage ensures that models evaluated on HoliDub-Bench are stress-tested across a representative spectrum of real-world recording conditions, ranging from controlled studio setups to challenging in-the-wild scenarios.

For background sound events, which are annotated as multi-label attributes to reflect the natural co-occurrence of multiple acoustic phenomena, "Electronic hum / equipment" is the most frequently occurring category (332 out of 1,000 samples), followed by "Echo / Reverberation" (246) and "Traffic / Road noise" (198). Other notable categories include "Footstep / Rustling" (117), "Crowd / Background voices" (113), and "Wind" (105), while less frequent events such as "Birds / Animal sounds" (21) and "Rain" (9) constitute a meaningful long tail. This rich, multi-label annotation scheme captures the layered complexity of real-world soundscapes and provides fine-grained diagnostic signals for analyzing model performance under diverse and compounding acoustic interferences.

\begin{table*}[!tbp]
\centering
\footnotesize
\caption{Holistic Generation vs. Decoupled Pipeline results on  HoliDub-Bench. }
\label{app_holibench}
\vspace{-6pt}
\resizebox{0.86\linewidth}{!}
{
\begin{tabular}{lccccccccccc}
\toprule
& Model & LSE-C$\uparrow$ & LSE-D$\downarrow$ & EMO-SIM$\uparrow$ & WER$\downarrow$ & UTMOS$\uparrow$ & MOS$\uparrow$ & FD$\downarrow$ & FAD$\downarrow$ & KLD$\downarrow$ & IS$\uparrow$ \\
\midrule
& GT & - & - & - & 14.51 & 2.80 & 3.89 & - & - & - & - \\
& HoliDubber (prompt) + AudioLDM & 5.21 & 8.82 & 0.980 & 16.67 & 2.03 & 3.81 & 17.65 & 5.53 & 2.01 & \textbf{1.65} \\
& HoliDubber   & \textbf{6.44} & \textbf{8.08} & \textbf{0.992} & \textbf{12.81} & \textbf{3.02} & \textbf{3.96} & \textbf{10.95} & \textbf{3.08} & \textbf{1.86} & 1.51 \\
\bottomrule
\end{tabular}
}
\end{table*}

\section{Audio Caption Prompt Design}
\label{app_pipeline}

\begin{figure}[htbp]
  \centering
  \includegraphics[width=\linewidth]{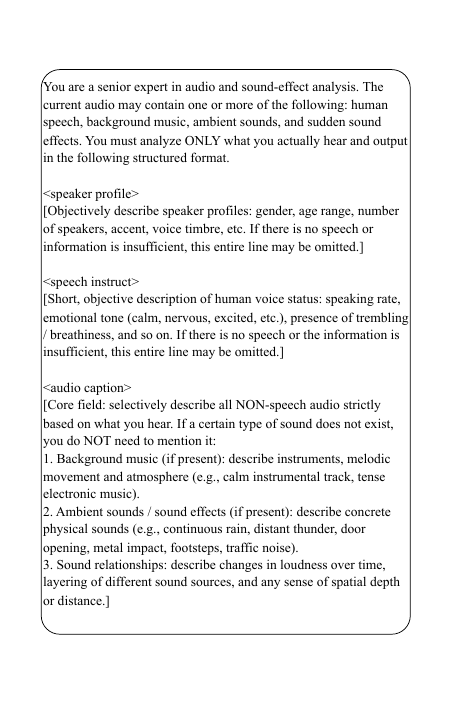}
  \caption{Overview of Prompt Design. }
  \label{audio_process_pipeline}
\end{figure}

To obtain structured and reliable audio annotations, we design a role-based prompt that instructs the captioning model to act as a senior expert in audio and sound-effect analysis. As illustrated in Figure 4, the prompt enforces a three-field output schema: \texttt{<speaker profile>}, \texttt{<speech instruct>}, and \texttt{<audio caption>}. The \texttt{<speaker profile>} field captures objective speaker attributes such as gender, estimated age range, accent, and voice timbre. The \texttt{<speech instruct>} field characterizes paralinguistic properties of the speech signal, including speaking rate, emotional tone, and vocal quality markers such as breathiness or trembling. Both fields are designed to be optional — the model is explicitly instructed to omit them when speech is absent or insufficient evidence is available, thereby minimizing hallucinated descriptions. The \texttt{<audio caption>} field serves as the core annotation target, requiring the model to selectively describe all non-speech audio content strictly grounded in what is actually heard. This field is further decomposed into three analytical dimensions: background music characterization (e.g., instrumentation, melodic contour, and atmosphere), ambient sounds and discrete sound effects (e.g., rain, footsteps, traffic noise), and inter-source relationships including temporal dynamics, loudness variations, layering, and spatial depth. By combining a grounded-only constraint with hierarchical structural guidance, this prompt design encourages comprehensive yet faithful audio descriptions while effectively suppressing speculative or fabricated content.

To ensure annotation reliability, we conduct a final round of human verification. Annotators review a stratified sample of the generated captions, checking for hallucinated sound events, misidentified speaker attributes, and inconsistencies between the caption and the actual audio content. Samples that fail quality checks are discarded, ensuring that the resulting annotations meet a consistent standard of accuracy before being used for model training and benchmark construction.

\section{Holistic Generation vs. Decoupled Pipeline}
\label{app_exp}

A natural alternative to our end-to-end approach is a decoupled pipeline that first synthesizes speech with a dedicated dubbing model, then separately generates background audio with a Text-to-Audio model, and finally mixes the two outputs. To evaluate this, we construct \textbf{HoliDubber (prompt) + AudioLDM}. Specifically, we run HoliDubber in the text-prompt-guided dubbing mode with the \texttt{<audio caption>} field removed from the structured text description, retaining only \texttt{<speaker profile>}, \texttt{<speech instruct>}, and \texttt{<text>} to generate speech-only output. In parallel, we feed the \texttt{<audio caption>} field alone to our pre-trained AudioLDM model to independently generate the corresponding background audio. The two outputs are then mixed at equal loudness to produce the final audio. We compare this decoupled pipeline against HoliDubber's holistic generation on HoliDub-Bench. 
To ensure semantic consistency with the pre-trained AudioLDM, we leverage Qwen-2.5 \cite{yang2024qwen25} to rewrite our the audio caption into the specific captioning style of the AudioCaps \cite{kim2019audiocaps} dataset.
% table

As shown in Table \ref{app_holibench}, HoliDubber consistently outperforms the decoupled pipeline across nearly all metrics. Most notably, the pipeline suffers a dramatic drop in UTMOS (2.03 vs 3.02), indicating that the post-mixing of independently generated speech and background audio significantly degrades overall audio quality. HoliDubber also achieves superior lip-sync accuracy (LSE-C: 6.44 vs 5.21), emotion similarity (0.992 vs 0.980), content accuracy (WER: 12.81 vs 16.67), and audio fidelity (FD: 10.95 vs 17.65, FAD: 3.08 vs 5.53). The only exception is IS, where the pipeline scores slightly higher (1.65 vs 1.51), likely because AudioLDM, as a dedicated audio generation model, produces more diverse acoustic patterns in isolation, however, this diversity comes at the cost of coherence with the speech content.

We attribute these results to three key advantages of end-to-end holistic generation. First, the decoupled pipeline treats speech and background audio as independent signals, inevitably introducing artifacts at the mixing stage—such as unnatural loudness transitions and temporal misalignment—that are directly reflected in the substantial UTMOS degradation. HoliDubber, by contrast, jointly models both components within a single generative process, producing naturally coherent audio where speech and background sounds are inherently blended. Second, in the decoupled pipeline, the AudioLDM model has no awareness of the speech content, and vice versa, leading to potential conflicts such as background music masking speech or ambient sounds being temporally inconsistent with the speaking dynamics. Our HoliDubber leverages shared contextual representations to coordinate speech and non-speech elements, as evidenced by the superior FD and FAD scores. Third, the pipeline approach requires separate inference passes for speech and audio generation followed by manual mixing, whereas HoliDubber generates the complete acoustic scene in a single forward pass, resulting in lower computational overhead and a simpler deployment workflow.

\section{Training Details}
\label{app_training_details}

\paragraph{\textbf{Audio-VAE}}
We train the Audio-VAE for 1120000 steps with a batch size of 128 on 8 NVIDIA H20 GPUs. We use the AdamW optimizer with an initial learning rate of 1e-4 and a exponential learning rate schedule. The total training data comprises approximately 30,000 hours. The audio is resampled to 32 kHz, and each training sample is cropped to 0.4 seconds. The latent dimension is set to 64. The training objective combines a reconstruction loss, a KL divergence term weighted by $\beta$=0.1, and a multi-scale discriminator adversarial loss following \cite{kumar2023dac}. The entire Audio-VAE training takes approximately 1200 GPU hours.

\paragraph{\textbf{Text-to-Audio Pre-training}}
We pre-train the text-guided audio generation backbone for 10 epochs with a batch size of 8000 vae latent frames on 32 NVIDIA H20 GPUs. The Causal Autoregressive Transformer is initialized from a pre-trained Qwen2.5-1.5B language model. We use the AdamW optimizer with a peak learning rate of 1e-4, a linear warmup of 20000 steps, and a constant schedule. 

\paragraph{\textbf{HoliDubber Training}}
We train HoliDubber on the combined training splits of VoxCeleb2 and CelebV-Dub for 50 epochs with a batch size of 8000 vae latent frames on 8 NVIDIA H20 GPUs. 
The Patch-based AV-Fusion module is randomly initialized, while other parameters are initialized from the pre-trained TTA backbone.
We use a constant learning rate of 8e-5  with the AdamW optimizer. The random text-prompt dropout probability is set to 0.5.

\end{document}